\documentclass{evn2004}
\usepackage{txfonts}
\usepackage{graphicx}

\begin{document}

\title{The asymmetric compact jet of GRS~1915+105}

\author{M. Rib\'o\inst{1}
\and V. Dhawan\inst{2}
\and I.~F. Mirabel\inst{1,3}
}

\institute{Service d'Astrophysique, CEA Saclay, B\^at. 709, L'Orme des Merisiers, 91191 Gif-sur-Yvette, Cedex, France
\and National Radio Astronomy Observatory, Socorro, NM 87801, USA
\and Instituto de Astronom\'{\i}a y F\'{\i}sica del Espacio, CONICET,
C.C.67, Suc. 28, 1428 Buenos Aires, Argentina
}

\abstract{We present multiepoch VLBA observations of the compact jet of
GRS~1915+105 conducted at 15.0 and 8.4~GHz during a {\it plateau} state of the
source in 2003 March-April. These observations show that the compact jet is
clearly asymmetric. Assuming an intrinsically symmetric continuous jet flow,
using Doppler boosting arguments and an angle to the line of sight of
$\theta=70\degr$, we obtain values for the velocity of the flow in the range
0.3--0.5$c$. These values are much higher than in previous observations of
such compact jet, although much lower than the highly relativistic values
found during individual ejection events. These preliminary results are
compatible with current ideas on the jet flow velocity for black holes in the
low/hard state.}

\maketitle

\section{Introduction}

GRS~1915+105 is a well-known microquasar containing a very massive
stellar-mass black hole of $M = 14 \pm 4~M_\odot$ (Greiner et~al.
\cite{greiner01}). During major flares it shows discrete relativistic ejection
events with a velocity in the range 0.90--0.98$c$, inferred from VLA and
MERLIN observations and assuming a distance to the source of $\simeq$12~kpc
(see Mirabel \& Rodr\'{\i}guez \cite{mirabel94} and Fender et~al.
\cite{fender99}, respectively). On the other hand, when the source is in the
so-called {\it plateau} state, a compact radio jet with a much lower velocity
of 0.1$c$ has been found in previous VLBA observations (Dhawan et~al.
\cite{dhawan00}). We observed the source with the VLBA of the
NRAO\footnote{The National Radio Astronomy Observatory is a facility of the
National Science Foundation operated under cooperative agreement by Associated
Universities, Inc.} during three epochs on spring 2003, as part of a
multiwavelength campaign presented in Fuchs et~al. (\cite{fuchs03}). Here we
present a preliminary analysis of the obtained VLBA images.

\section{Observations and results}

We observed \object{GRS~1915+105} with the VLBA at 15.0 and 8.4~GHz on 2003
March 24, April 2 and April 19 (runs A, B and C, respectively). A summary of
the time intervals of our observations is presented in Table~\ref{summary}.
The source \object{1923+210} was used as fringe-finder, while
\object{J1924+154} was used for phase-referencing the observations of
\object{GRS~1915+105} and the check source \object{J1922+084}. Cycle times of
3 minutes were used at both frequencies. The observations were conducted at
256 Mb~s$^{-1}$, with 8 baseband channels and 2 bit sampling using both
polarizations (modes {\tt v2cm-256-8-2.set} and {\tt v4cm-256-8-2.set}). This
provided tape passes of 22 minutes at each frequency, that included a scan on
the fringe-finder, 5 cycles on the target source and 1 cycle on the check
source. We alternated 15.0 and 8.4~GHz passes, and we could include a total of
8, 6 and 11 passes on runs A, B and C, respectively. The data were processed
at the VLBA correlator in Socorro, with an integration time of 2~s.

\begin{figure}
\resizebox{\hsize}{!}{\includegraphics{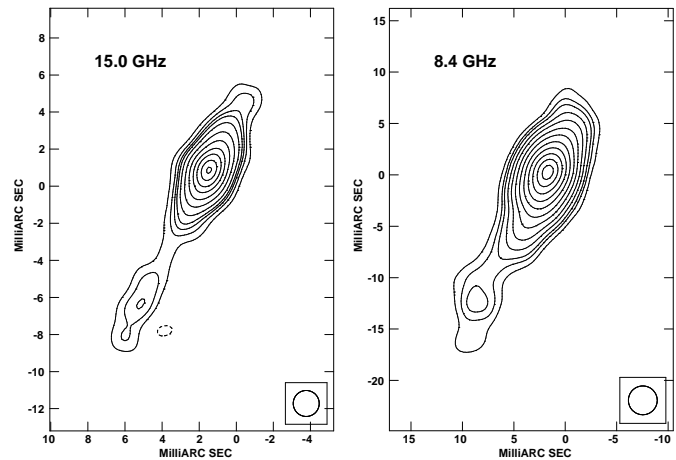}}
\caption[]{VLBA images at 15.0 and 8.4~GHz obtained on 2003 April 19.
The rms noise of the images are 0.19 and 0.08~mJy~beam$^{-1}$, respectively, and contour levels start with $-$5, 5, 10, 15 times the rms noise.
Notice the different scale used in each image.
}
\label{vlba}
\end{figure}

The data reduction was performed using standard procedures within the {\sc
aips} software package of NRAO. We used the following parameters to produce
the images: a weighting value for robust of $-$1, a cellsize of 0.15 and
0.30~mas, and a $uv$-tapering of 120 and 60~k$\lambda$ at 15.0 and 8.4~GHz,
respectively. Several phase self-calibration steps using boxes along the jet
direction were performed, as well as careful editing of bad data after the
first iterations. We produced final images by using a circular beam with a
Full Width at Half Maximum (FWHM) equal to the average of the geometric means
of the FWHM of the beams at all epochs: 1.4 and 2.8~mas at 15.0 and 8.4~GHz,
respectively. All the obtained images reveal a jet-like feature, with a
slightly asymmetric compact component brighter to the southeast (the
approaching part of the jet) plus a further elongation towards the southeast.
As an example, we show in Fig.~\ref{vlba} the images obtained at both
frequencies during run C.

\begin{table*}
\begin{center}
\caption[]{Summary of the VLBA observations. See text for a detailed explanation.}
\label{summary}
\begin{tabular}{c@{~~}c@{~~~}c@{~~}c@{~~~}c@{~~~}c@{~~~}c@{~~~}c@{~~~~~}c@{~~~~}c@{~~~~~}c@{~~~~}c}
\hline \hline \noalign{\smallskip}
Run   & MJD     & UT                        & $\nu$ & $S_{\nu}$ & $\alpha$     & $S_{\nu ~ \rm app}$ & $S_{\nu ~ \rm rec}$ & $\beta\cos\theta$ & $\beta$     & $\beta\cos\theta$ & $\beta$ \\
      &         &                           & (GHz) & (mJy)     &              & (mJy)               & (mJy)               & \multicolumn{2}{c}{(for $k=2$)} & \multicolumn{2}{c}{(for $k=3$)} \\
\noalign{\smallskip} \hline \noalign{\smallskip}

A     & 52722.7 & 2003 Mar 24~~15:30--18:30 & 15.0  & 131.4     & +0.3$\pm$0.2 & 80.1                & 55.1                & 0.11$\pm$0.02   & 0.32$\pm$0.04 & 0.07$\pm$0.01 & 0.20$\pm$0.02 \\
      &         &                           &~~8.4  & 108.3     &              & 63.2                & 46.0                & 0.09$\pm$0.01   & 0.27$\pm$0.03 & 0.06$\pm$0.01 & 0.17$\pm$0.01 \\

B     & 52731.6 & 2003 Apr 02~~13:00--15:00 & 15.0  & 122.7     & +0.1$\pm$0.2 & 83.9                & 41.4                & 0.18$\pm$0.02   & 0.54$\pm$0.06 & 0.12$\pm$0.01 & 0.35$\pm$0.03 \\
      &         &                           &~~8.4  & 115.8     &              & 63.5                & 53.3                & 0.05$\pm$0.01   & 0.13$\pm$0.02 & 0.03$\pm$0.01 & 0.09$\pm$0.01 \\

C     & 52748.4 & 2003 Apr 19~~08:30--12:40 & 15.0  & 112.0     & +0.4$\pm$0.2 & 69.6                & 43.8                & 0.14$\pm$0.02   & 0.42$\pm$0.05 & 0.09$\pm$0.01 & 0.26$\pm$0.02 \\
      &         &                           &~~8.4  &~~88.4     &              & 55.2                & 33.5                & 0.15$\pm$0.02   & 0.45$\pm$0.06 & 0.10$\pm$0.01 & 0.28$\pm$0.02 \\
      
\noalign{\smallskip} \hline
\end{tabular}
\end{center}
\end{table*}

We also conducted VLA observations of \object{GRS~1915+105} at 22 and 8.4~GHz
during the first hour of run A, at 8.4~GHz during the first hour of run B, and
again at 22 and 8.4~GHz just one day before run C. The obtained lightcurves
show a nearly constant flux density during each individual observation,
consistent with the steady behavior found in both Ryle Telescope radio
observations and {\it RXTE}/ASM X-ray count rate (see Fig.~2 in Fuchs et~al.
\cite{fuchs03}). Therefore, the requirement of constant flux density for
synthesis imaging is satisfied, but see a detailed discussion on the eventual
problems related to a changing morphology during the observations in Dhawan
et~al. (\cite{dhawan00}).

To better study the asymmetry of the jet we proceeded in the following way.
First we selected the Clean Components (CC) representing the jet on the final
images, and exported them out of {\sc aips}. Then we shifted their coordinates
to have the maximum centered at (0,0). We finally rotated them by using the
Position Angle (PA) of the jet, 155\degr, so the X coordinate of the CC is
parallel to the jet direction defined positive towards the approaching jet,
and the Y coordinate is perpendicular to it. We then added together the flux
of all CC in the approaching jet, and we did the same for the receding one.

Assuming that the jet is intrinsically symmetric, and neglecting possible
free-free absorption effects, we can obtain the product $\beta\cos\theta$,
being $\beta$ the velocity of the flow in units of the speed of light and
$\theta$ the angle between the jet and the line of sight, by means of the
equation:
\begin{equation}
\beta\cos\theta={\big({S_{\nu ~ \rm app}/{S_{\nu \rm ~ rec}}}\big)^{1/(k-\alpha)}-1 \over \big({S_{\nu ~ \rm app}/{S_{\nu \rm ~ rec}}}\big)^{1/(k-\alpha)}+1}~,
\label{eqflux}
\end{equation}
where $S_{\nu ~ \rm app}$ and $S_{\nu ~ \rm rec}$ are the flux densities of
the approaching and receding jets, respectively, $k$ equals 2 for a continuous
jet and 3 for discrete condensations, and $\alpha$ is the spectral index
defined as $S_{\nu}\propto \nu^{+\alpha}$. Since the equation above is only
valid for fluxes measured at equal distances from the core, we have added to
the flux density obtained with the CC for the receding jet, the quantity
$n\times3\sigma$, being $\sigma$ the rms noise in the image and $n$ the result
of dividing the difference in total distance of the approaching and receding
jets by the FWHM of the beam.

We quote in Table~\ref{summary} the obtained results, where $S_\nu$ is the
total flux density of the jet, $\alpha$ has been computed from the $S_\nu$
values at both frequencies on a given epoch (we assume a standard error of
0.2), and $S_{\nu ~ \rm app}$ and $S_{\nu ~ \rm rec}$ correspond to the fluxes
of the approaching and receding jets computed as explained above. We finally
quote the derived values for $\beta\cos\theta$ for the continuous jet ($k=2$)
and for the less realistic case of discrete condensations ($k=3$), together
with the value of $\beta$ derived by assuming an angle $\theta=70\degr$ in
each case (see Fender et~al. \cite{fender99}). Values within the quoted errors
have been obtained by changing the PA of the jet by $\pm5\degr$. Since the
images reveal a continuous jet, we will only consider the $k=2$ case. The
obtained values of $\beta$ are $\simeq0.3$ for run A and $\simeq0.4$ for run
C, while very different values are obtained at different frequencies for run
B, being around 0.5 at 15.0~GHz and around the much lower value of 0.1 at
8.4~GHz. This last result can be understood when inspecting the corresponding
image because the maximum CC happens to be slightly towards the receding part
of the jet, and only a small shift of 0.3~mas in the jet center would provide
a value of $\beta=0.38\pm0.04$.

\section{Discussion and conclusions}

In general we can say that with the procedure described above we obtain a
value for the jet flow velocity of $\sim0.4~c$ when assuming $k=2$ and
$\theta=70\degr$. Lower values of $\beta$ are obtained for smaller angles
$\theta$, which should be used if the source is closer than 12~kpc (Fender
et~al. \cite{fender99}; Chapuis \& Corbel \cite{chapuis04}). In any case, the
obtained values are higher than the ones reported in Dhawan et~al.
(\cite{dhawan00}) for a similar compact jet, but still much lower than the
highly relativistic values inferred in discrete ejection events. These
preliminary results are compatible with the Lorentz factors of $\Gamma\la2$
allowed for compact radio jets if the empirical correlation found between
X-ray/radio flux for black holes in the low/hard sate holds (Gallo et~al.
\cite{gallo03}). A detailed analysis of these observations and their
implications will be presented in a forthcoming paper.

\begin{acknowledgements}
M.R. acknowledges support by a Marie Curie Fellowship of the European
Community programme Improving Human Potential under contract number
HPMF-CT-2002-02053, partial support by DGI of the Ministerio de Ciencia
y Tecnolog\'{\i}a (Spain) under grant AYA2001-3092, as well as partial support
by the European Regional Development Fund (ERDF/FEDER).
\end{acknowledgements}

\end{document}